\documentclass{iaus}
\usepackage{graphicx}

  \checkfont{eurm10}
  \iffontfound
    \IfFileExists{upmath.sty}
      {\typeout{^^JFound AMS Euler Roman fonts on the system,
                   using the 'upmath' package.^^J}%
       \usepackage{upmath}}
      {\typeout{^^JFound AMS Euler Roman fonts on the system, but you
                   dont seem to have the}%
       \typeout{'upmath' package installed. iaus.cls can take advantage
                 of these fonts,^^Jif you use 'upmath' package.^^J}%
      }
  \else
  \fi

  \checkfont{msam10}
  \iffontfound
    \IfFileExists{amssymb.sty}
      {\typeout{^^JFound AMS Symbol fonts on the system, using the
                'amssymb' package.^^J}%
       \usepackage{amssymb}%

      }{}
  \fi

  \IfFileExists{amsbsy.sty}
    {\typeout{^^JFound the 'amsbsy' package on the system, using it.^^J}%
     \usepackage{amsbsy}}
    {\providecommand\boldsymbol[1]{\mbox{\boldmath $##1$}}}

\newsavebox{\astrutbox}
\sbox{\astrutbox}{\rule[-5pt]{0pt}{20pt}}

\title[Non-thermal phenomena in galaxy clusters]
      {Non-thermal phenomena in galaxy clusters}

\author[G.Brunetti]
{Gianfranco Brunetti$^{1}$}
\affiliation{$^1$ Istituto di Radioastronomia INAF/CNR,
via P.Gobetti, 101, I--40129 Bologna, Italy} 

\pubyear{2004}
\volume{195}
\pagerange{1--8}
\date{?? and in revised form ??}
\jname{Outskirts of Galaxy Clusters: intense life in the suburbs}
\editors{A. Diaferio, ed.}

\begin{document}
\maketitle

\begin{abstract}
The discovery of diffuse synchrotron radio emission and, more
recently, of the
hard X-ray (HXR) tails have triggered a growing interest about  
non-thermal phenomena in galaxy clusters.
After a brief review of the most important evidences for non-thermal
emission, I will focus on the origin of the
emitting particles and of the hadronic component.
In particular I will describe the particle-injection and -acceleration 
mechanisms at work in the intra-cluster medium (ICM) and, at the
same time, discuss the possibility to test current modellings 
of these phenomena with future radio, HXR, and gamma rays observatories.
\end{abstract}

\section{Introduction}

There is now firm evidence that the ICM is a mixture of hot gas, magnetic
fields and relativistic particles. 
While the hot gas results in thermal
bremsstrahlung X-ray emission, relativistic electrons
and positrons generate non-thermal radio (synchrotron) and 
hard X-ray radiation (inverse Compton). 
In principle, the amount of the
energy budget of the intracluster medium in the form of high energy
hadrons can be large, due to the confinement of cosmic rays
over cosmological time scales (V\"{o}lk et al. 1996;
Berezinsky, Blasi \& Ptuskin 1997).
The collisions between thermal and relativistic hadrons
in the ICM generate $\pi^o$ and secondary $e^{\pm}$. 
Both these species radiate a relevant fraction of their 
energy into the gamma band 
via $\pi^o$ decay and via $e^{\pm}$-inverse Compton scatter of the 
photons of the cosmic microwave background.
However, such gamma radiation that would allow us to
constrain the energetics of relativistic hadrons in clusters 
has not been detected as yet (Reimer et al., 2003).

\begin{figure}
\includegraphics[width=6.5cm]{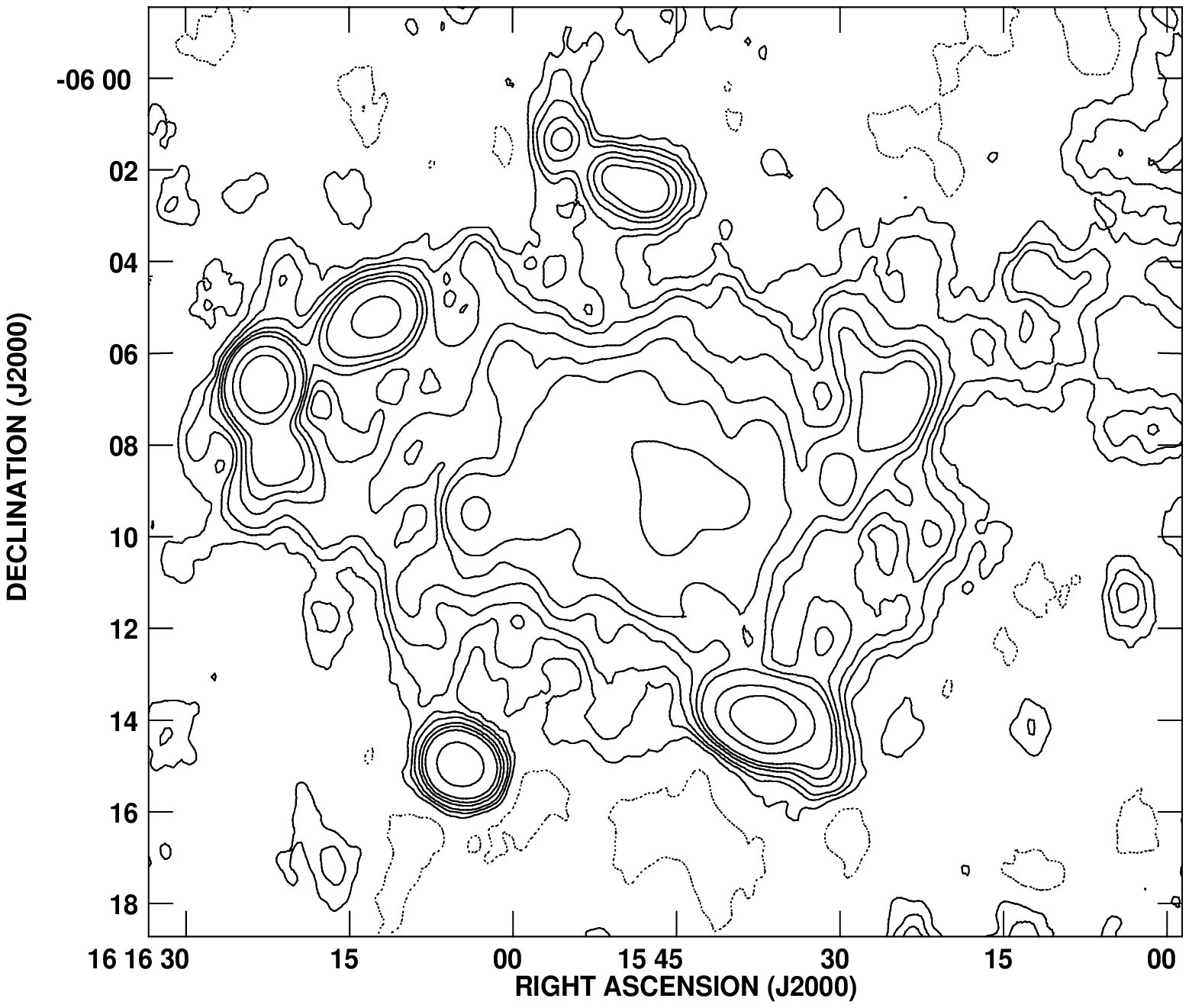}
\hfill
\includegraphics[width=6.0cm]{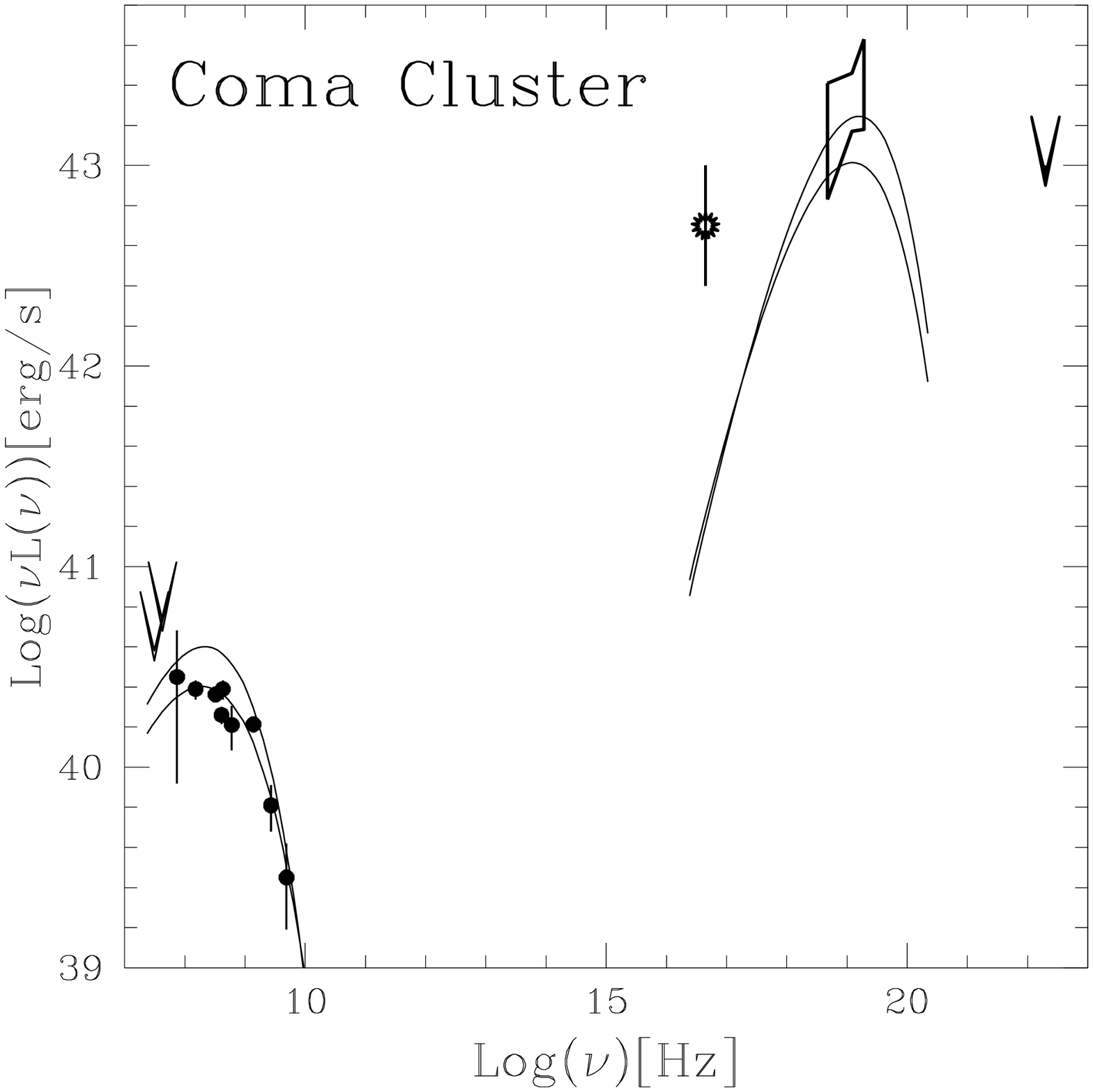}
\caption{\small
{\sc Left Panel}: 327 MHz image of the giant radio halo in A 2163
(Feretti et al.2004). 
{\sc Right Panel}: Broad band non-thermal spectrum of the Coma radio halo:
the fit to the radio and HXR data is provided by synchrotron and IC emission 
from a reaccelerated electron population. 
}
\end{figure}

\section{Observations}

The most important evidence for relativistic electrons in clusters
of galaxies comes from the diffuse synchrotron radio emission observed in
a growing number of massive clusters
(e.g., Feretti, 2003).
The diffuse emissions are referred to as {\it radio halos} 
(Fig.1a) and/or radio
{\it mini--halos} when they appear confined to the center of the
cluster, while they are called {\it radio relics} when they are found in the
cluster periphery.
Giovannini, Tordi and Feretti (1999) found that $\sim$5\% of clusters
in a complete X-ray flux limited sample have a diffuse radio source.
The detection rate of diffuse radio sources shows a abrupt increase
with the X-ray luminosity of the host clusters:
about 30-35\% of the galaxy clusters with
X-ray luminosity larger than 10$^{45}$ erg s$^{-1}$
show diffuse radio emission
(Feretti, 2003).
Interestingly, there is a correlation between the non--thermal
diffuse radio emission and the presence of merger activity in
the host clusters of galaxies (Buote, 2001; Schuecker et al.2001):
this suggests a link between the process of formation of
galaxy clusters and the origin of the non-thermal activity.

A second important evidence for relativistic electrons
comes from the hard X-ray (HXR) excess emissions detected in
a few galaxy clusters by the BeppoSAX and RXTE satellites
(Fusco-Femiano et al.2003a; Rephaeli et al.2003).
HXRs may be explained in terms of IC scattering of relativistic
electrons off the photons of the cosmic microwave background 
(e.g., Fusco--Femiano et al.2003a)
and thus they represents a unique tool to unambiguously disentangle
between the energy of the relativistic electrons and of the 
magnetic field in the ICM.
Unfortunately the poor sensitivity of the present
and past HXR-facilities does not allow to obtain a 
iron-clad detection of HXR excesses
and thus future observatories (e.g. ASTRO-E2, NEXT) are necessary 
to definitely confirm these excesses (see Rossetti \& Molendi 2004;
Fusco-Femiano et al.2004).

Additional evidence for non-thermal emission may also 
come from the extreme ultra-violet (EUV) excesses
discovered in a number of galaxy clusters
(e.g., Bowyer 2003).
EUV excesses, and their origin, are however still matter of debate
(see however Bonamente and Kaastra, these proceedings).

In Fig.1b a compilation of the non-thermal spectrum 
of the Coma cluster is reported.

\section{Modelling the origin of the emitting leptons}
So far relativistic leptons are the unique non-thermal 
component detected in the ICM and thus our present understanding of the
non-thermal activity is mostly
based on the modelling of this component.
The most spectacular example of non-thermal emission in
galaxy clusters is given by the giant {\it radio halos} (Fig.1a).
The difficulty in explaining these sources arises from the
combination of their $\sim$Mpc size, and the relatively short radiative
lifetime of the radio emitting electrons (about $10^8$yrs):
the diffusion time
necessary to these electrons to cover such distances is orders of
magnitude larger than their radiative lifetime.
Thus the emitting electrons cannot be  
injected by clusters' Galaxies and/or AGN and then simply 
diffuse into the ICM.

\begin{itemize}
\item[{\it i)}]{\it primary models}:
as proposed first by Jaffe (1977), a solution to this puzzle would be
provided by continuous {\it in situ} reacceleration of the relativistic
electrons on their way out; this possibility was then studied more 
quantitatively by Schlickeiser et al.(1987).

\item[{\it ii)}]{\it secondary models}:
an alternative to the {\it in situ} 
reacceleration scenario was put forward by
Dennison (1980), who suggested that relativistic electrons may be
continuously 
injected in the ICM by inelastic proton-proton collisions through
production and decay of charged pions.
\end{itemize}

\subsection{Broad band spectrum}
A first possibility to constrain the spectrum and origin
of relativistic leptons is given by the study of the 
broad band spectrum of the non-thermal emission (e.g., Fig.1b).

\begin{itemize}
\item[{\it i)}]{\it HXRs and $\gamma$-Rays}:
If the HXR excess in the Coma cluster is of IC origin, 
then the number of relativistic electrons in the ICM should be
large and this can be combined with the EGRET upper limit to reject
the hypothesis of a secondary origin of the emitting
particles. Indeed in this case a large number of relativistic hadrons 
is required and the gamma ray flux produced via $\pi^o$ decay 
should overproduce the EGRET upper limit (e.g., Blasi \& Colafrancesco 1999).
Thus {\it secondary models} should admit a different
origin for the HXRs. One possibility is given
by supra-thermal bremsstrahlung emission
(Blasi 2000; Dogiel 2000) which however would 
require a too large energy budget to maintain the HXR excess 
for more than $10^8$yrs (Petrosian, 2001).

\item[{\it ii)}]{\it EUV and HXR Excesses}:
A second possibility is given by the 
combination of the EUV and HXR excesses.
If HXRs are due to IC scattering by the same electrons
which emit the synchrotron radiation,
then a low frequency (e.g., $< 10$ keV) flattening of the photon 
spectrum is required.
The presence of this flattening indicates 
a corresponding flattening of the spectrum of the electrons at the energies 
responsible for IC emission below the HXR band which 
can result from Coulomb 
losses if the emitting electrons are in the cluster core,
or it would be also naturally obtained if particles are
accelerated by stochastic processes (e.g., 
via turbulent acceleration, Fig.1b).
\end{itemize}

It is clear that the HXR fluxes are pivot points 
in the modelling of the non-thermal spectra
of galaxy clusters and thus that the advent of the
future observatories (ASTRO-E2, NEXT) will be crucial to confirm
the above issues.

\subsection{Detailed Properties of the Radio Emission}

Additional possibilities to constrain the origin of
the emitting particles derive from the study of the 
detailed radio properties of a few well studies {\it radio halos}:

\begin{itemize}

\item[{\it i)}]{\it Broad synchrotron radio profiles}:
Giant {\it radio halos} are very extended sources with 
an extension up to 2-2.5 Mpc. 
The radial synchrotron profiles of these giant {\it radio halos} 
are found to be
broader than that of the X-rays emitted by the hot gas (e.g., Govoni
et al.2001).
This basically means that the synchrotron emissivity 
($j \propto K_e B^{1+\alpha}$ for a spectrum of electrons 
$N(\gamma) = K_e \gamma^{-(1+2\alpha)}$) decreases with distance 
from the cluster center less rapidly than the bremsstrahlung emissivity 
$j \propto n_{th}^2$ ($n_{th}$ is the density of the thermal plasma),
and thus that the spatial distribution of the relativistic
electrons is broader than that of the thermal particles.
If radio electrons are of {\it secondary origin} then
a very large number of relativistic hadrons is required in the 
clusters outskirts (since the production rate of secondary particles 
depends on $n_{th}$) and this causes a serious energetic problem 
at least if $\mu$G central fields and a relatively steep decrease 
of the field strength from the cluster center 
(as theoretically expected, e.g. Dolag et al.2002) 
are assumed (Brunetti 2003).
Obviously the energetic problem may be alleviated in some cases
(e.g. Coma) by assuming stronger central fields and a slower 
radial decrease 
of the field strength (Pfrommer \& Ensslin 2004).

\item[{\it ii)}]{\it Observed synchrotron spectra}:
The integrated synchrotron spectrum of a few
{\it radio halos} steepens at high frequencies (e.g., Giovannini et al.1993;
Fusco-Femiano et al.2003b).
Although, in some cases, the observed steepening may be mitigated by 
taking into account the 
SZ effect (Ensslin, 2002), these 
observations point out to the presence of a high energy
break or a cut-off in the spectrum of the emitting electrons.
A second point is that the 0.3-1.4 GHz spectral index maps 
of an increasing number of {\it radio halos} indicate a 
progressive steepening of the radio spectrum from
the center to the periphery of the clusters (Fig.2a);
the radial steepening in the Coma cluster, originally
discovered with interferometric observations, is also 
required by single dish observations (Deiss et al., 1997).
Radial spectral steepenings are theoretically
explained in terms of the presence
of a high energy break or cut-off in the electron spectrum 
combined with a radial decrease of the ICM-field strength 
(Brunetti et al.2001; Kuo et al.2003).
Such high energy cut-offs should be at energy 
$\gamma \sim 10^3-10^5$
and are naturally expected if the electrons are 
re-accelerated by some kind of mechanism (e.g., Brunetti et al.2001; 
Petrosian 2001).

\item[{\it iii)}]{\it Synchrotron spectral and brightness variations}:
A more recent observational hint
is that most {\it radio halos} show synchrotron brightness increments 
and complex spectral features in the spectral-index maps 
in coincidence with dynamically disturbed regions
of the clusters and with temperature patches (Markevitch et al.2002; 
Feretti et al.2004).
These observations suggest a close link between the dynamical
status of the ICM and the spectrum of the particles (and/or
the magnetic field) on 100-200 kpc scales.
\end{itemize}

Although future observations are obviously required to confirm 
all the above findings, it is clear that 
when the broad radial profiles of {\it radio halos} 
are {\it consistently} combined with the observed 
radial spectral steepenings, then {\it secondary models}
have inescapable problems.

\section{Cluster Mergers \& Particle Acceleration: a general view}

In this Section we release the ``historical''
dichotomy between {\it primary}
and {\it secondary} origin of the emitting particles and  
give a more general (but brief) theoretical overview of the mechanisms 
which are believed to drive non-thermal processes;
a simple scheme is given in Fig.2b. 

\subsection{Injection Processes}
The first step in our understanding of non-thermal activity 
is given by the modelling of the injection and of the spatial diffusion
of relativistic particles in the
ICM (Fig.2b, bottom-left),
and of the energy released by cluster mergers into 
shock waves and turbulence which will
be responsible for particle acceleration (Fig.2b, top-center).

\begin{itemize}
\item[{\it i)}]{\it Injection of relativistic particles in the ICM}:
clusters contains Galaxies and AGN which would
inject relativistic leptons and hadrons which will remain
confined in the cluster volume (V\"{o}lk et al. 1996;
Berezinsky et al.1997).
We don't know which is the fraction of energy channeled by these 
sources into relativistic leptons and hadrons, 
relativistic leptons, however, radiate their energy in a time-scale
much shorter than a Hubble time, and thus it is very likely that 
the non-thermal energy stored in galaxy clusters 
is dominated by the hadronic component.
Relativistic hadrons in the ICM continuously generate secondary leptons 
due to collisions with thermal protons and these
particles contribute to the population of relativistic leptons
in the ICM.

\item[{\it ii)}]{\it Cluster Mergers: injection of Shocks and Turbulence}:
accretion of matter at the virial radius is likely to 
form strong shocks, while cluster mergers 
should drive shock waves in the internal
regions of the clusters (e.g., Miniati et al.2000 and ref. therein).
There is still some debate on the typical Mach number of the shocks developed
during mergers. 
Some numerical simulations suggest that a relatively
large fraction of these shocks 
have a high Mach number (Miniati et al., 2000, 2001). 
On the other, hand semi--analytical
calculations (Gabici \& Blasi 2003; Berrington \& Dermer 2003) 
find Mach numbers 
of order unity as also observed by {\it Chandra} 
(e.g., Markevitch et al.2003).
More recent numerical simulations (Ryu et al., 2003) seem 
to find more weak shocks than in Miniati et al., 
however, the comparison with analytical calculations appears difficult 
because of a different classification of the shocks in the two approaches.

Fluid turbulence is expected to be injected in galaxy clusters
during cluster mergers. Numerical simulations find that an energy
budget of 10-30\% of the thermal energy of the ICM 
can be associated to cluster turbulence (Sunyaev et al.2003); 
this may be tested directly with future
X-ray observatories (ASTRO-E2).
In addition, first evidence for cluster turbulence has been found  
by the analysis of 
a very recent Newton-XMM observation of the Coma cluster (Schuecker et
al.2004).
\end{itemize}

\subsection{Particle Acceleration}
The second point related to the generation of non-thermal phenomena is 
the interplay between particles and acceleration mechanisms.
The idea is that this interplay is particularly efficient
during cluster merger events.
Two mechanism of interplay are reported in Fig.2b: 
shock-particle coupling ({\it shock loop}, Fig.2b bottom-center) and
turbulence-particle coupling ({\it turbulence loop}, Fig.2b center-right).

\begin{itemize}
\item[{\it i)}]{\it Shock Loop}:
Collisionless shocks are generally recognized as efficient particle
accelerators through the so-called ``diffusive shock acceleration'' (DSA) 
process (Blandford \& Eichler 1987).
This mechanism has been invoked several times as acceleration
process in clusters of galaxies (Takizawa \& Naito, 2000; Blasi 2001;
Miniati et al., 2001; Fujita \& Sarazin 2001). 
Besides the scale and morphology of the 
synchrotron emission of {\it radio halos} is not naturally
accounted for by shock-accelerated particles, 
evidence for shock acceleration in galaxy clusters
may come from the so called {\it radio relics} which indeed
can be produced through the combination of shock acceleration 
and adiabatic compression of {\it ghost radio plasma} by shock 
waves propagating into the ICM (e.g., Ensslin 2003).
\begin{figure}
\includegraphics[width=5.5cm]{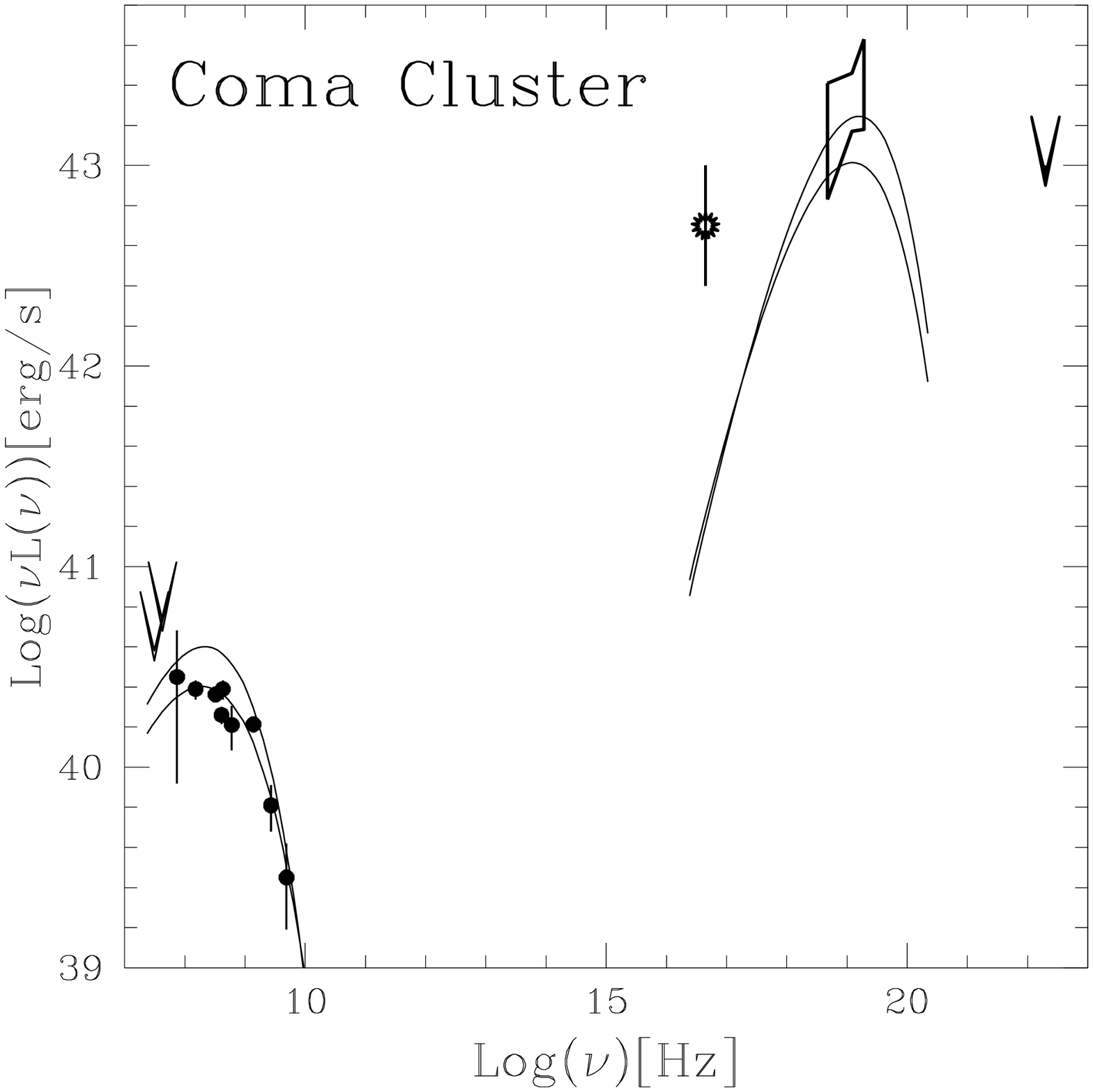}
\hfill
\includegraphics[width=7.0cm]{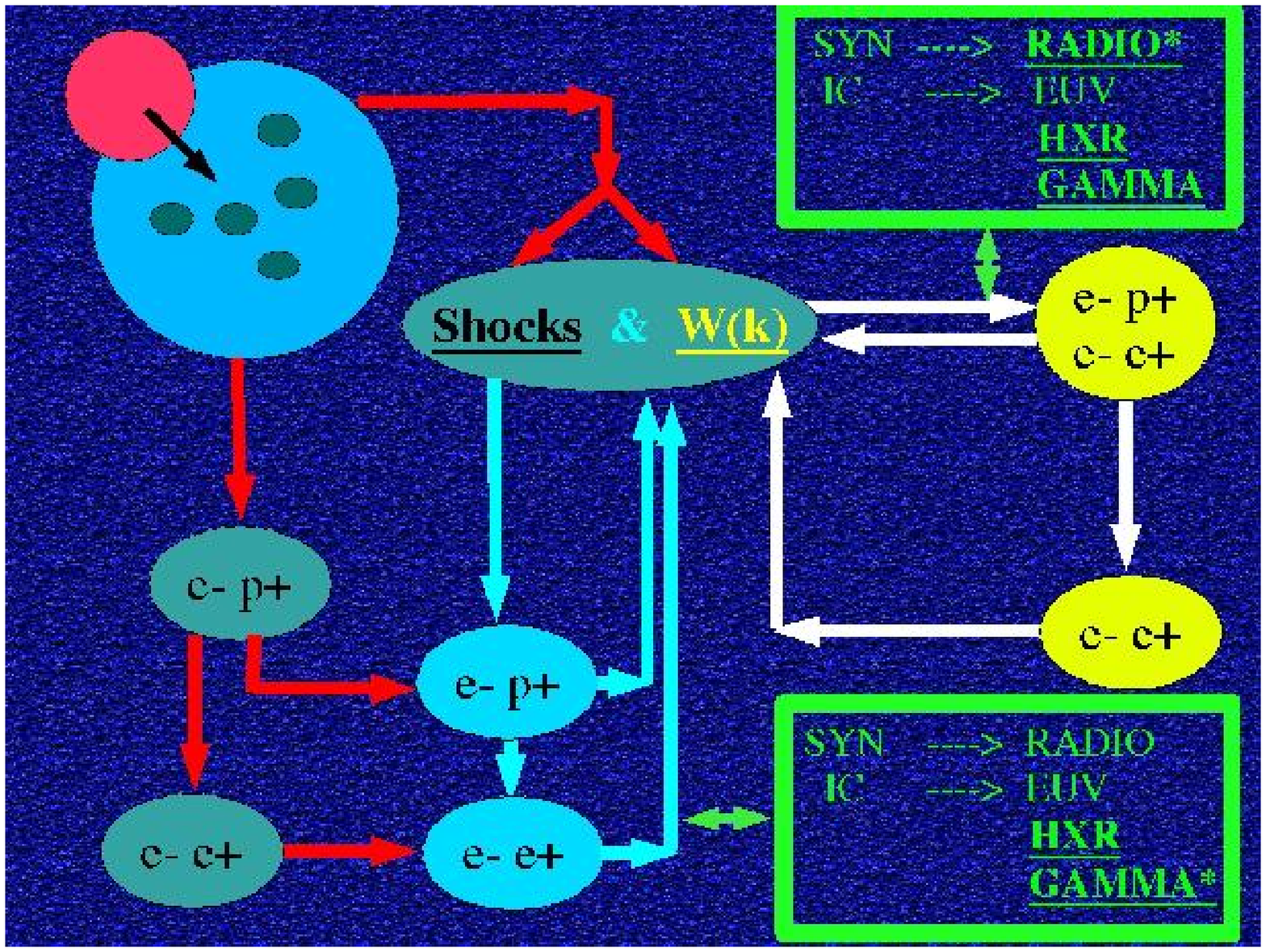}
\caption{\small
{\sc Left Panel}: Spectral index (327-1400 MHz) with distance from the
cluster center for Coma (boxes, Giovannini et al.1993), A665 (filled points) 
and A2163 (empty points, Feretti et al.2004).
{\sc Right Panel}: 
Scheme of the processes in galaxy clusters (see text).
}
\end{figure}
Shocks in the ICM may accelerate particles from the thermal pool and/or
reaccelerate seed relativistic particles in the ICM with a power law
in momentum with a slope 
$= 2( {\cal M}^2 +1 )/( {\cal M}^2 -1) [+1]$ (${\cal M}$ is the 
Mach number and ``[+1]'' should be added to calculate the 
spectrum of particles in the ICM under the assumption of a stationary
continuous particle-injection by shock waves).
If shocks related to major mergers are weak as suggested by
semi-analytical calculations (Sect.4.1), they accelerate very
steep particle-spectra and thus they are not the dominant
process for the injection of relativistic particles in the ICM.
However, the maximum energy of particles accelerated at a shock wave is 
very high:

\begin{equation}
E_{\it max}^e ({\it GeV})\sim
2.5 \times 10^4 B_{\mu G}^{1/2} v_8 
\,\,\,\,\,\,\,\,\,\,\,\,\,\,\,\,\,
{\it and}\,\,\,\,\,\,\,\,\,\,\,\,\,\,\,\,\, 
E_{\it max}^p ({\it GeV})\sim
4.5 \times 10^8 B_{\mu G} v_8^2 
\end{equation}

for leptons and hadrons, respectively
($v_8$ is the speed of the shock in units of $10^8$km/s, $B$ is
the field in units of $\mu$G and Bohm diffusion coefficient
is assumed).
The decay of $\pi^o$ generated in hadronic collisions together
with the IC emission from {\it secondary} leptons
injected by the same collisions and from 
{\it primary} electrons accelerated at shock waves 
will produce gamma ray radiation above 100 MeV (e.g., Blasi 2001).
The amount of radiation depends on the energy
budget of relativistic hadrons and on their spectrum.
Gamma rays emitted from the outskirts of
galaxy clusters are expected to be dominated by IC emission from
primary electrons accelerated at accretion shocks, these 
shocks are indeed expected to be very strong and thus the spectrum
of the accelerated electrons should be flat ($\propto E^{-2}$).
On the other hand, given the different results on the Mach number of 
merger shocks obtained by different approaches (Sect.4.1), 
there is no general consent on the gamma ray
flux and spectrum expected from the central regions of galaxy clusters
where the contributions from the 
products of hadronic collisions should be important
(e.g., Miniati 2002).
Hopefully future gamma ray observations 
will unambiguously clarify this issue and will allow us to constrain
the energetics and spectrum of relativistic hadrons in galaxy clusters
(Gabici \& Blasi 2004).

\item[{\it ii)}]{\it Turbulence Loop}:
The origin of the giant {\it radio halos} and possibly of the
HXR tails is most likely associated with the turbulence loop.
Indeed, it has been shown that re--acceleration of a population of relic 
electrons by turbulence powered by major mergers is suitable to explain 
the very large scale of the observed radio emission and is also a promising 
possibility to account for the complex spectral behaviour observed in the
diffuse radio sources and for the HXR tails (Brunetti et al., 2001a,b; 
Petrosian 2001; Ohno, Takizawa and Shibata 2002; 
Fujita, Takizawa and Sarazin 2003).

\item[{\it iii)}]{\it Self-Consistent Modellings}:
Very recently, the problem of particle-Alfv\'en wave interactions has
been investigated in the most general situation in which relativistic 
electrons, thermal protons and relativistic protons exist within the 
cluster volume (Brunetti et al., 2004). 
In this modelling the interaction of all these components with the waves, 
as well as the turbulent cascading and damping processes of Alfv\'en waves, 
have been treated in a fully time-dependent way in order to 
calculate the spectra of electrons, protons and waves at any fixed time. 
This work has provided a first investigation of the importance of the presence 
of the relativistic protons in the ICM for the electron acceleration.
The most important result of this work is that Alfv\'enic electron
acceleration can produce the observed phenomena provided that 
relativistic protons are not dynamically important in the ICM 
(less than few percents of the thermal energy).
Of course additional MHD waves which do not interact with particles
via resonant-acceleration (e.g., Magnetosonic waves) may increase the
efficiency of the electron acceleration without being very
sensitive to the presence of relativistic hadrons.
In addition, if reaccelerated, the spectrum of the hadronic component 
becomes harder and this may increase the efficiency in producing secondary 
particles in the ICM.
These particles can be reaccelerated interacting
with the MHD waves as in the case of the primary seed electrons,
so that the leptonic acceleration can efficiently generate 
non-thermal phenomena also in the case in which the number
of primary leptons in the ICM is negligible (Brunetti \& Blasi, in prep).
\end{itemize}

\acknowledgments
We acknowledge  partial financial support from INAF (Istituto
Nazionale di Astrofisica) through grant D4/03/IS.
It is a pleasure to thank my collaborators and colleagues
P.Blasi, R.Cassano, L.Feretti, S.Gabici and G.Setti for
useful discussions.

\end{document}